\documentstyle[prl,eqsecnum,aps,floats]{revtex} 
\begin{document}

\title{Observation of the Cabibbo--suppressed decay 
$\Xi_{c}^{+} \rightarrow pK^-\pi^+$}

\author{                                                                      
S.Y.~Jun,$^{3}$ 
N.~Akchurin,$^{16}$ 
V.~A.~Andreev,$^{11}$ 
A.G.~Atamantchouk,$^{11}$ 
M.~Aykac,$^{16}$ 
M.Y.~Balatz,$^{8}$ 
N.F.~Bondar,$^{11}$ 
A.~Bravar,$^{20}$ 
P.S.~Cooper,$^{5}$ 
L.J.~Dauwe,$^{17}$
% ...
\\                                                                            
\vskip 0.50cm                                                                 
\centerline{(SELEX Collaboration)}                                             
\vskip 0.50cm                                                                 
}                                                                             
\address{  
\centerline{$^{1}$Ball State University, Muncie, Indiana 47306}
\centerline{$^{2}$Bogazici University, Bebek 80815 Istanbul, Turkey}
\centerline{$^{3}$Carnegie--Mellon University, Pittsburgh, Pittsburgh 15213}
\centerline{$^{4}$Centro Brasileiro de Pesquisas F\'{\i}sicas, Rio de Janeiro,
                  Brazil}
\centerline{$^{5}$Fermi National Accelerator Laboratory, Batavia, 
                  Illinois 60510}
\centerline{$^{6}$Institute for High Energy Physics, Protvino, Russia}
\centerline{$^{7}$Institute of High Energy Physics, Beijing, PR China}
\centerline{$^{8}$Institute of Theoretical and Experimental Physics, 
                  Moscow, Russia}
\centerline{$^{9}$Max--Planck--Institut f\"ur Kernphysik, 69117 Heidelberg,
                  Germany}
\centerline{$^{10}$Moscow State University, Moscow, Russia}
\centerline{$^{11}$Petersburg Nuclear Physics Institute, St. Petersburg, 
                   Russia}
\centerline{$^{12}$Tel Aviv University, 69978 Ramat Aviv, Israel}
\centerline{$^{13}$Universidad Aut\'onoma de San Luis Potos\'{\i}, 
                   San Luis Potos\'{\i}, Mexico}
\centerline{$^{14}$Universidade Federal da Para\'{\i}ba, Para\'{\i}ba, Brazil}
\centerline{$^{15}$University of Bristol, Bristol BS8~1TL, United Kingdom}
\centerline{$^{16}$University of Iowa,  Iowa City, Iowa  52242}
\centerline{$^{17}$University of Michigan--Flint, Flint, Michigan 48502}
\centerline{$^{18}$University of Rome "La Sapienza" and INFN , Rome, Italy}
\centerline{$^{19}$University of S\~ao Paulo, S\~ao Paulo, Brazil}
\centerline{$^{20}$University of Trieste and INFN, Trieste, Italy}
}
\date{\today}
\maketitle

\begin{abstract}

We report the first observation of the Cabibbo--suppressed 
charm baryon decay $\Xi_{c}^{+} \rightarrow pK^{-}\pi^{+}$.  
We observe $150 \pm 22$ events for the signal.
The data were accumulated using the SELEX spectrometer during the 1996--1997 
fixed target run at Fermilab, chiefly from a 600 GeV/$c$ $\Sigma^{-}$ beam.  
The branching fractions of the decay relative to the Cabibbo--favored
$\Xi_{c}^{+} \rightarrow \Sigma^{+}K^{-}\pi^{+}$ and 
$\Xi_{c}^{+} \rightarrow \Xi^{-}\pi^{+}\pi^{+}$ are measured to be
$B(\Xi_{c}^{+} \rightarrow pK^{-}\pi^{+}) /B(\Xi_{c}^{+} \rightarrow 
\Sigma^{+}K^{-}\pi^{+}) = 0.22 \pm 0.06 \pm 0.03$ and
$B(\Xi_{c}^{+} \rightarrow pK^{-}\pi^{+}) /B(\Xi_{c}^{+} \rightarrow 
\Xi^{-}\pi^{+}\pi^{+}) = 0.20 \pm 0.04 \pm 0.02$, respectively.
\end{abstract}


\twocolumn
\input{psfig}

The study of Cabibbo--suppressed (CS) charm decays can provide useful insights 
into the weak interaction mechanism for non--leptonic decays~\cite{Korner}.
The observed final state may arise either from direct quark emission at the
decay stage or, in some cases, from quark rearrangement due to final--state
scattering.  By comparing the strengths of CS decays to their
Cabibbo--favored (CF) analogs, one can, in a systematic way, assess the 
contributions of the various mechanisms. 

Modern methods for calculating non--leptonic decay rates of the charm hadrons 
employ heavy quark effective theory and the factorization
approximation~\cite{HQEH}.  Nonetheless, the three--body decays of charm 
baryons are prohibitively difficult to calculate due to the complexity of 
associated final state interactions.  Measurements of the relative branching 
fractions of charm baryon states, both CF and CS, give additional information
about the structure of the decay amplitude and the validity of the 
factorization approximation.  

Many CS hadronic decays of charm mesons have been measured by both $e^{+}e^{-}$
collider and fixed target experiments.  Until now the only CS charm baryon 
decay reported with significant statistics 
is $\Lambda_{c}^{+} \rightarrow pK^{-}K^{+}$ and its resonant state,
$\Lambda_{c}^{+} \rightarrow p\phi$~\cite{CLEO-1,E687-1}.  This transition
requires internal W--emission and may be inhibited by color--suppression
effects.  In this paper, we present the first observation of the CS decay 
$\Xi_{c}^{+} \rightarrow pK^{-}\pi^{+}$ and determine the branching fractions 
of this decay relative to the CF
$\Xi_{c}^{+} \rightarrow \Sigma^{+}K^{-}\pi^{+}$ and 
$\Xi_{c}^{+} \rightarrow \Xi^{-}\pi^{+}\pi^{+}$ modes.  All three of these
modes involve external W--emission.

\begin{figure}[h]
\centerline{\psfig{figure=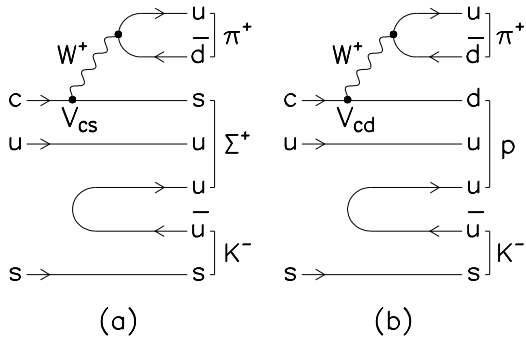,width=80mm}}
\caption{An example of spectator diagrams for $\Xi_{c}^{+}$ decays, 
(a) CF $\Xi_{c}^{+} \rightarrow \Sigma^{+}K^{-}\pi^{+}$ and 
(b) CS $\Xi_{c}^{+} \rightarrow pK^{-}\pi^{+}$}
\label{fig:xcp decay}
\end{figure}

Figure \ref{fig:xcp decay} shows a spectator diagram with external
$W$--emission for $\Xi_{c}^{+}$ decaying into a CF and a CS mode.
(The other CF $\Xi^-$ mode interchanges $s$ and $u$ quark lines and produces
a $d\overline{d}$ pair from the vacuum instead of a $u\overline{u}$ pair.)    
The decay process is similar in the two modes except for the flavor change in
the weak decay Cabibbo--Kobayashi--Maskawa matrix 
element ($V_{cs}$ vs. $V_{cd}$).  Based on a simplified dimensional 
analysis, we would expect 
$B(\Xi_{c}^{+} \rightarrow pK^{-}\pi^{+})/
 B(\Xi_{c}^{+} \rightarrow \Sigma^{+}K^{-}\pi^{+}) = 
 2.1 \times \alpha \times {\rm tan}^{2}\theta_{c}$
where $\theta_{c}$ is the Cabibbo angle and $\alpha$, of order unity, 
measures the ratio of the hadronic contributions to the weak decay matrix
element.  The factor 2.1 reflects the relative phase space of the two modes.
By comparing the $\alpha$ factors for the entire family of CS non--leptonic
charm baryon decays one may develop a picture of the dominant features of the
amplitudes.

The SELEX(E781) experiment at Fermilab is particularly well--suited for 
studying CS charm baryon decays because of its excellent particle 
identification capabilities.  SELEX is a high energy hadroproduction 
experiment using a 3--stage spectrometer designed for high acceptance for
forward interactions ($x_F = 2p_{\parallel}^{CM}/\sqrt{s} > 0.1$). 
The experiment emphasizes understanding charm production in the forward
hemisphere and the study of charm baryon decays.  Using both a negative  
beam ($\simeq 50$\% $\Sigma^-$, $\simeq 50$\% $\pi^-$) and positive 
beam ($\simeq 92$\% $p$, $\simeq 8$\% $\pi^+$), SELEX recorded 15.2 billion 
interaction events during the 1996--1997 fixed target run.  The majority
particles ($\Sigma^-$, $\pi^-$) in the 600 GeV/$c$ negative beam and the
protons in the 540 GeV/$c$ positive beam were tagged by a beam 
transition radiation detector.  The data were accumulated from a segmented
target (5 foils: 2 Cu, 3 C, each separated by 1.5 cm) whose total 
thickness was  5\% of an interaction length for protons.

The spectrometer had silicon strip detectors to measure the beam and outgoing
tracks, giving precision primary and secondary vertex reconstruction.
Transverse position resolution ($\sigma$) was 4 $\mu{\rm m}$ for the 600 
GeV/$c$ beam tracks.  The average longitudinal vertex position resolution was 
270 $\mu{\rm m}$ for primary and 560 $\mu{\rm m}$  for secondaries, 
respectively.  Momenta of particles deflected by the analyzing magnets were 
measured by a system of proportional wire chambers (PWCs), drift chambers and
silicon strip detectors.  Momentum resolution for a typical 100 GeV/$c$ track
was $\sigma_p/p \approx 0.5$\%.  The absolute momentum scale was calibrated
using the $K_S^0$ mass.  For $D^0 \rightarrow K^- \pi^+$ decays the average
mass resolution was 9 MeV/$c^2$ independent of $D^0$ momenta 
from 100--450 GeV/$c$.  Charged particle identification was done with a Ring
Imaging $\check{\rm C}$erenkov detector (RICH)~\cite{rich-nim}, which 
distinguished $K^{\pm}$ from $\pi^{\pm}$ up to 165 GeV/$c$.  The proton 
identification efficiency was greater than 95\% above proton threshold
($\approx 90$ GeV/$c$).  For pions the total mis--identification probability
due to all sources of confusion was less than 4\%.

Interactions were selected by a scintillator trigger.  The trigger for charm
required at least 4 charged tracks after the targets as indicated by an
interaction counter and at least 2 hits in a scintillator hodoscope after the
second analyzing magnet.  It accepted about 1/3 of all inelastic interactions.
Triggered events were further tested in an on--line computational filter 
based on downstream tracking and particle identification information.
The on--line filter selected events that had evidence of a secondary vertex
from tracks completely reconstructed using the forward PWC spectrometer and
the vertex silicon.  This filter reduced the 
data size by a factor of nearly 8 at a cost of about a factor of 2 in 
charm written to tape (as determined from a study of 
unfiltered $K_S^0$, $\Lambda$ and $D^0 \rightarrow K^-\pi^+ + c.c.$ decays).
Most of the charm loss came from selection
cuts that were independent of charm species or kinematic variables and
which improved the signal/noise in the final sample.

Results presented here come from the first processing pass through all data.
In this analysis, secondary vertex reconstruction was attempted when 
the $\chi^2$ per degree of freedom ($\chi^2$/dof) for the fit of the ensemble
of tracks to a single primary vertex exceeded 5.  All combinations of tracks
were formed for secondary vertices with as many as 5 prongs.  Secondary 
vertices were tested against a reconstruction table that specified 
selection criteria for each charm decay mode.
Secondary vertices which occurred inside the volume of a target or after the 
first plane of the silicon vertex detector were rejected.

The $pK^{-}\pi^{+}$ sample was selected by the following requirements:
(i) primary and secondary vertex fits each have $\chi^{2}/{\rm dof} < 5$; 
(ii) vertex separation significance $L/\sigma \ge 8$, where $L$ is the
longitudinal separation between primary 
and secondary vertices and $\sigma$ is the error on $L$;  
(iii) the reconstructed momentum vector from the secondary vertex points
back to the primary vertex with  $\chi^{2}/{\rm dof} < 4$;
(iv) the momentum of the $\pi^{+}$ $\ge$ 5 GeV/$c$;
(v) $\mathcal{L}$($K$)/$\mathcal{L}(\pi) > {\rm 1}$ for $K$ identification 
and $\mathcal{L}$($p$)/$\mathcal{L}(\pi) > {\rm 1}$ for $p$ identification, 
where $\mathcal{L}$ is the mass--selection likelihood function for the RICH;
and (vi) the sum of the squared
transverse momenta of the three daughter particles with respect to the parent
direction, $\Sigma p_{T}^2 > 0.3$ ${\rm (GeV/}c)^2$. 

\begin{figure}[h]
\centerline{\psfig{figure=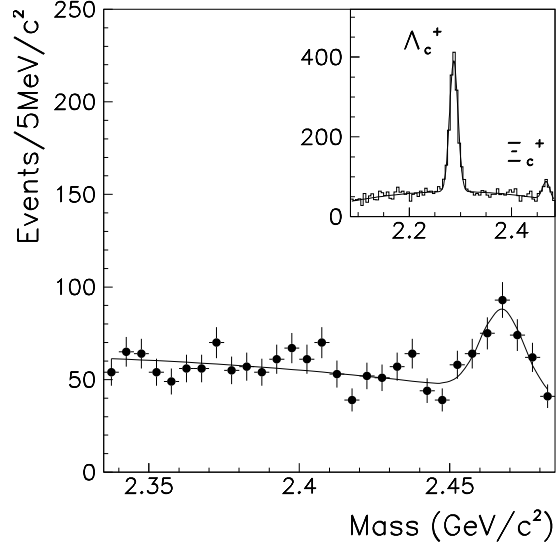,width=80mm}}
\caption{The invariant mass distribution of $pK^-\pi^+$.}
\label{fig cspki}
\end{figure}

Figure \ref{fig cspki} shows the first observation of the CS
$\Xi_{c}^+ \rightarrow p K^-\pi^+$ decay mode.  The inset of the figure shows 
the invariant mass distribution from the entire selected mass range of 
reconstructed $p K^-\pi^+$ candidates.  The large peak in the inset is the 
decay $\Lambda_{c}^+ \rightarrow pK^-\pi^+$.  The bump at the right 
is the CS $\Xi_{c}^+ \rightarrow pK^-\pi^+$ decay.
The distribution has an artificial high--mass cutoff because of a maximum 
mass cut imposed in this first--pass analysis.
Monte Carlo simulations show that the observed number of signal events
is systematically reduced only by ($2 \pm 3$)\% due to this cutoff.
The background characteristics are very similar both in the number per mass bin
and lifetime through the entire mass region, so that the 
background subtraction 
under the $\Xi_c^+$ peak is stable.  The lifetime both of background and 
signal events for the $\Xi_c^+$ region has been investigated.  The background 
apparent lifetime is quite short, while the $\Xi_c^+$ lifetime is consistent
with the Particle Data Group average~\cite{PDG}.
A detailed lifetime analysis is underway and will be reported elsewhere.
The mass is determined using a Gaussian of variable centroid, width and area.
The number of signal events is determined by a sideband subtraction method.
The signal window is determined from the $\Lambda_c^+$ data and is set at
45 MeV/$c^2$.  The background is a second order 
polynomial fit to the entire mass region excluding a 65 MeV/$c^2$ wide 
window centered at each mass peak.  The $\Xi_{c}^+$ yield is the difference 
between the summed events in the 45 MeV/$c^2$ signal window centered 
at 2.467 GeV/$c^2$ and the extrapolated background.  Variations
in the exclusion window size change the extrapolated background by fewer
than 2 events.
There are $150 \pm 22$ events for $\Xi_{c}^+ \rightarrow pK^-\pi^+$ at a 
$\Xi_{c}^+$ mass of $2467.4 \pm 1.3$ MeV/$c^2$.  The statistical significance 
for the signal, $S/\sqrt{S+B}$, is $6.9 \pm 0.6$ in which $S$ is the number of 
signal events and $B$ is the number of 
background events in the signal region. 

Background contributions from 
charm baryon ($\Xi_{c}^{0}$ and $\Omega_{c}^{0}$) decay to the 
CS decay are expected to be negligible; they produce an even
number of secondary vertex tracks and their decay lengths are short 
compared to the minimum $\Xi_{c}^+$ flight path cut.  
Excellent proton identification strongly suppresses background from 
charm meson decays (three--body $D^+$ and $D_{s}^+$).

\begin{figure}[h]
\centerline{\psfig{figure=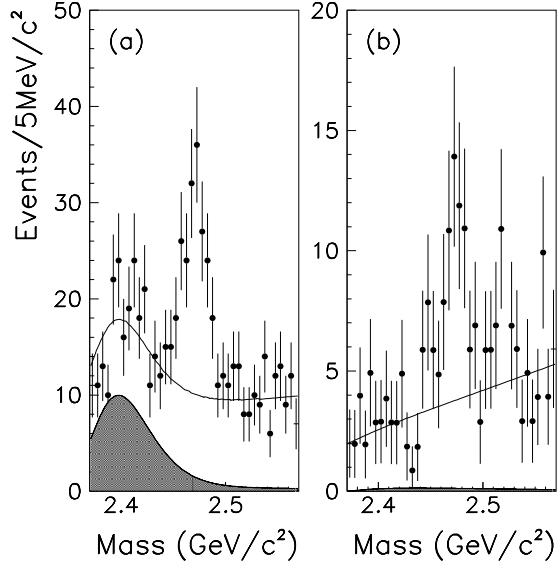,width=80mm}}
\caption{Charm baryon $\Xi_{c}^+$ signals with hyperon partial 
reconstruction, (a) $\Xi^-\pi^+\pi^+$ and (b) $\Sigma^+K^-\pi^+$.
The solid line is the background fit.}
\label{fig xc+}
\end{figure}

One normalizes the CS mode by comparing it to CF modes that have a hyperon 
in the final state.  In the first reconstruction pass,
hyperons ($\Sigma^{\pm}, \Xi^{-}$) were identified only inclusively in a
limited decay interval (5 -- 12 m downstream from the target): candidates 
had $p > 40$ GeV/$c$ and no hits assigned along the trajectory in the 14 
chambers after the second analyzing magnet.  This category of tracks gives 
unique $\Sigma^+$ identification but is ambiguous between
$\Sigma^{-}$ and $\Xi^{-}$.  Figure \ref{fig xc+} shows two CF $\Xi_{c}^+$
modes decaying to $\Xi^-\pi^+\pi^+$ and $\Sigma^+K^-\pi^+$, respectively.  
The shaded areas in Fig. 3(a) and 3(b) are the
estimated reflection from $\Lambda_{c}^+ \rightarrow \Sigma^-\pi^+\pi^+$
and $\Lambda_{c}^+ \rightarrow \Sigma^+\pi^-\pi^+$, respectively.  
The shapes are determined by a Monte Carlo simulation and the areas are 
normalized to the observed number of signal events in the $\Lambda_c^+$ data.
For this part of the analysis, additional cuts were applied to all data.
The pion momentum threshold was raised from 5 to 10 GeV/$c$ and the
transverse component of the $\Xi_{c}^+$ momentum with respect to its line of 
flight was required to be less than 0.3 GeV/$c$.  

The total acceptance (geometrical acceptance and reconstruction efficiency)
for decay modes of interest was estimated by embedding Monte Carlo charm decay
tracks into data events.  Momentum and energy were not conserved in the 
process, but studies indicate this has little effect on the single--charm
acceptance calculation.  Events were generated with an average transverse 
momentum $\langle p_{T} \rangle = 1.0$ GeV/$c$ and longitudinal momentum 
distributions as observed for the data.  Detector hits, including resolution
and multiple Coulomb scattering smearing effects, produced
by these embedded tracks were folded into the hit banks of the underlying data 
event.  The new ensemble of hits was passed through the SELEX off--line 
software.  The acceptance is the ratio of the number of events in the signal
to the number of embedded events in a particular mode.

The method was checked by measuring two well--determined relative branching
fractions,  
$B(D^0 \rightarrow K^-\pi^+\pi^+\pi^-)/
B(D^0 \rightarrow K^-\pi^+) = 1.94 \pm 0.07 \pm^{0.09}_{0.11}$ and 
$B(D^{+} \rightarrow K^-K^+\pi^+)/
B(D^{+} \rightarrow K^-\pi^+\pi^+) = 0.093 \pm 0.010 \pm^{0.008}_{0.006}$
where the first error is statistical and the second is systematic.
Systematic errors include the difference of the ratio for charge--conjugate 
states, and uncertainties of signal yields and acceptance estimations.  
The results agree well with the world 
average~\cite{PDG}.  For the $D^{+}$ decays, the vertex significance cut
($L/\sigma$) was increased to 20 to suppress background from 
$D_{s}^{+} \rightarrow K^-K^+\pi^+$.  This costs little signal, since the 
lifetime of $D^{+}$ is 2.4 times longer than that of $D_{s}^{+}$.

Using this procedure, we measure the relative branching
fraction for 
$\Xi_{c}^{+} \rightarrow \Sigma^{+}K^{-}\pi^{+}$ and 
$\Xi_{c}^{+} \rightarrow \Xi^{-}\pi^{+}\pi^{+}$ to be
$(0.92 \pm 0.20 \pm 0.07)$.  
The systematic error is due to
uncertainties in the background subtraction and relative acceptance 
estimation. This result is comparable to the CLEO 
measurement, $(1.18 \pm 0.26 \pm 0.17)$~\cite{CLEO-3}.  
The number of events and estimated acceptance for the three observed 
$\Xi_{c}^+$ modes with the common set of cuts are summarized in 
table~\ref{tab xc+}.

\begin{table}
\centering
\begin{tabular}{|l||r|r|}
                                        & Acceptance(\%)   & Events~~ \\ \hline \hline
$\Xi_{c}^+ \rightarrow \Xi^-\pi^+\pi^+$ & $2.46 \pm 0.04$ & $130 \pm 15$ \\ \hline
$\Xi_{c}^+ \rightarrow \Sigma^+K^-\pi^+$& $1.09 \pm 0.02$ & $ 53 \pm 10$ \\ \hline
$\Xi_{c}^+ \rightarrow pK^-\pi^+$       & $7.08 \pm 0.06$ & $ 76 \pm 13$ 
\end{tabular}
\caption{Summary of observed events and estimated acceptance 
for $\Xi_{c}^+$ modes using the tighter cuts. The errors are statistical only.}
\label{tab xc+}
\end{table}

The branching fraction of the CS decay
$\Xi_{c}^{+} \rightarrow pK^{-}\pi^{+}$ relative to the CF
$\Xi_{c}^{+} \rightarrow \Sigma^{+}K^{-}\pi^{+}$ is measured to be
$0.22 \pm 0.06 \pm 0.03$ which corresponds to
$(4.2 \pm 1.1 \pm 0.5) \times {\rm tan}^{2}\theta_{c}$.
The $\alpha$ factor is $2.0 \pm 0.5 \pm 0.2$.  To put this in
some perspective, the corresponding $\alpha$ factor for the
CS decay $\Lambda_c^+ \rightarrow pK^-K^+$ relative to the CF
mode $\Lambda_c^+ \rightarrow pK^-\pi^+$ is $2.5 \pm 0.6$, using the world
average for the branching fraction and correcting for phase space.
Thus, the factors for these two different baryons are similar.
In contrast, the CS ratios of $D$ mesons depend strongly on final state
multiplicity, suggestive of sizable final state interactions.
For the $\Xi^{-}\pi^{+}\pi^{+}$ mode, 
B($\Xi_{c}^{+} \rightarrow pK^{-}\pi^{+}$)/
B($\Xi_{c}^{+} \rightarrow \Xi^{-}\pi^{+}\pi^{+}$) is measured to be 
$0.20 \pm 0.04 \pm 0.02$.
Systematic errors include uncertainties in (i) the relative acceptance 
estimation, (ii) background subtractions due to the reflection from 
other hyperon modes, and (iii) the $pK^-\pi^+$ yield due to the 
imposed mass cutoff. 

In summary, we observe the CS decay 
$\Xi_{c}^{+} \rightarrow pK^{-}\pi^{+}$ at a mass of $2467.4 \pm 1.3$ MeV/$c^2$
with $150 \pm 22$ signal events.  The relative branching fractions of the decay
$\Xi_{c}^{+} \rightarrow pK^{-}\pi^{+}$ to the CF
$\Xi_{c}^{+} \rightarrow \Sigma^{+}K^{-}\pi^{+}$ and
$\Xi_{c}^{+} \rightarrow \Xi^{-}\pi^{+}\pi^{+}$
are measured to be
$B(\Xi_{c}^{+} \rightarrow pK^{-}\pi^{+}) /B(\Xi_{c}^{+} \rightarrow 
\Sigma^{+}K^{-}\pi^{+}) = 0.22 \pm 0.06 \pm 0.03$ and 
$B(\Xi_{c}^{+} \rightarrow pK^{-}\pi^{+}) /B(\Xi_{c}^{+} \rightarrow 
\Xi^{-}\pi^{+}\pi^{+}) = 0.20 \pm 0.04 \pm 0.02$, respectively.

%
The authors are indebted to the staffs of Fermi National Accelerator 
Laboratory, the Max--Planck--Institut f\"ur Kernphysik, Carnegie Mellon 
University, and Petersburg Nuclear Physics Institute for invaluable 
technical support.  
This project was supported in part by Bundesministerium f\"ur Bildung, 
Wissenschaft, Forschung und Technologie, Consejo Nacional de 
Ciencia y Tecnolog\'{\i}a {\nobreak (CONACyT)},
Conselho Nacional de Desenvolvimento Cient\'{\i}fico e Tecnol\'ogico,
Fondo de Apoyo a la Investigaci\'on (UASLP),
Funda\c{c}\~ao de Amparo \`a Pesquisa do Estado de S\~ao Paulo (FAPESP),
the Israel Science Foundation founded by the Israel Academy of Sciences and 
Humanities, Istituto Nazionale de Fisica Nucleare (INFN),
the International Science Foundation (ISF),
the National Science Foundation (Phy \#9602178),
NATO (grant CR6.941058-1360/94),
the Russian Academy of Science,
the Russian Ministry of Science and Technology,
the Turkish Scientific and Technological Research Board (T\"{U}B\.ITAK),
the U.S. Department of Energy (DOE grant DE-FG02-91ER40664 and DOE contract
number DE-AC02-76CHO3000), and
the U.S.-Israel Binational Science Foundation (BSF).
We also acknowledge a useful conversation with Austin Napier.

\end{document}